\documentclass[authoryear,12pt]{article}
\usepackage[english]{babel}
\usepackage{url,amsfonts,epsfig}
\usepackage{amsfonts}
\usepackage{stmaryrd}
\usepackage{mathrsfs}
\usepackage{amsmath}
\usepackage{enumerate}
\usepackage{mathrsfs}
\usepackage{mathtools}
\usepackage{graphicx}
\usepackage{amsthm}
\usepackage{natbib}
\usepackage{subfigure}
\usepackage{vmargin}
\usepackage{amssymb}
\usepackage{tikz}
\usetikzlibrary{plotmarks}
\newcommand\marksymbol[2]{\tikz[#2,scale=1.8]\pgfuseplotmark{#1};}

\setpapersize{A4}

\setpapersize{A4}
\setmarginsrb{23mm}{23mm}{23mm}{30mm}{0pt}{0mm}{0pt}{0mm}
\begin{document}
\title{On The Inverse Geostatistical Problem of Inference on Missing Locations}
\author{Emanuele Giorgi \& Peter J. Diggle \\
             Lancaster Medical School \\
             Lancaster University}
\maketitle
\begin{abstract}
The standard geostatistical problem is to predict the values of a spatially continuous phenomenon, 
$S(x)$ say, at locations $x$ using data $(y_i,x_i):i=1,..,n$ where  $y_i$  is the
realisation at location $x_i$ of $S(x_i)$, or of a random variable $Y_i$ that is stochastically
related to $S(x_i)$. 
In this paper we address the inverse problem of predicting the locations of
observed measurements $y$. We discuss how knowledge of the sampling mechanism can and should
 inform  a prior specification, $\pi(x)$ say, for the joint distribution of the
measurement locations $X = \{x_i: i=1,...,n\}$,
and propose an efficient Metropolis-Hastings algorithm for drawing samples from the
resulting predictive distribution of the missing elements of $X$.
An important feature in many applied settings is that this predictive
distribution is multi-modal, which severely limits the usefulness of simple summary measures
such as the mean or median. We present two simulated examples to
demonstrate the importance of the specification for $\pi(x)$, and
analyse rainfall data from Paran\'a State, Brazil
to show how, under additional assumptions,  an
empirical of estimate of $\pi(x)$ can be used when no prior information
on the sampling design is available.
 \vspace{0.4cm}
\end{abstract}

\section{Introduction}
\label{sec:intro}
Geostatistics was originally developed as a self-contained methodology for spatial prediction
 (e.g. \citet{matheron1963}) but is now embedded as a sub-branch of spatial statistics
with applications in many different disciplines. The canonical geostatistical problem
is to predict the value of a spatially continuous process, $S(x)$ say, at any required
location $x$ in a region of interest $A \subset \mathbb{R}^2$, using data consisting of a set of measured values $y_i$ at each of $n$ locations $x_i$ in $A$. A
widely used geostatistical model is that the $y_i$ are realisations of random variables
$Y_{i} = S(x_{i})+Z_{i}$, where $Z_{i}$ are mutually independent, zero-mean
Gaussian variables, and $S=\{S(x): x \in \mathbb{R}^2\}$ is a Gaussian process \citep{diggle1998}. 
Predictive inference for $S$ is then 
based on the predictive distribution $[S|Y]$, where $[\cdot]$ means ``the distribution of''  and
$Y=(Y_1,...,Y_n)$. Conventionally, the 
set of measurement locations $X=(x_1,...,x_n)$ is regarded either 
as fixed or, if stochastic, as 
non-informative in the sense that $S$ and $X$ are stochastically independent; 
\citet{diggle2010} refer to this assumption as {\it non-preferential} sampling.

In this paper we address the inverse problem of predictive inference for missing locations
associated with a subset of the measured variables $Y$. Our interest in this
problem was motivated by \citet{wasser2004}. They proposed a geostatistical model 
for data on 399 DNA samples of elephant tusks collected from 28 distinct known 
locations acrossAfrica, and showed how
their model could be used to infer the geographic location of a DNA sample of unknown origin,
in order to identify areas of intense poaching.  
For their Bayesian analysis, they specified a uniform prior 
for sample locations over either savannah or forest regions of Africa,
according to each sample's known origin and used
the corresponding posterior median of each spatial coordinate as a point prediction of the unknown 
location. They evaluated predictive performance empirically using
leave-one-out cross-validation, 
 and found that  their geostatistical approach 
outperformed  other methods that do not use geographic information. 
\par
Our objectives in this paper are the following: to extend the approach of \citet{wasser2004} 
 to include multiple missing locations;  to show how prior information on the sampling 
design can and should inform the analysis; and to highlight 
the unsatisfactory nature of simple statistics
such as the mean or the median as summaries of the 
predictive distribution of missing locations.\par 

The paper is structured as follows. In Section \ref{sec:model} we propose a modelling framework
 for predictive inference on multiple missing locations, incorporating prior knowledge
about the sampling design. 
 In Section \ref{sec:conditional_simulation} we show how
numerical quadrature can be used for predictive inference on a single unknown location,
and propose an MCMC algorithm for sampling from the 
joint predictive distributions of multiple unknown locations. In both cases we assume
that parameter values are known, which in practice corresponds to ignoring the uncertainty
parameter estimates. In Section \ref{sec:examples}, we
describe two simulated examples to illustrate the limitations
of simple summary statistics of the predictive distribution for the missing locations.  
We also report an analysis of rainfall data from Para\'na State, Brazil, where 
we use a non-parametric density estimate as an empirical prior for $X$.
 Section \ref{sec:discussion} is a concluding discussion. 

\section{Model formulation}
\label{sec:model}

\subsection{Measurement data}

We adopt a standard geostatistical model,
\begin{equation}
Y_i = \mu_i + S(x_i) + Z_i: i=1,...,n,
\label{eq:linear}
\end{equation}
where $S=\{S(x): x \in \mathbb{R}^2\}$ is
 a zero-mean stationary Gaussian process with variance $\sigma^2$ and correlation function $\rho(x,x';\phi)$ indexed by the parameter $\phi$, and the $Z_i$ are mutually independent
${\rm N}(0,\tau^2)$ variates. Equivalently, the $Y_i$ are conditionally independent given
$\{S(x): x \in \mathbb{R}^2\}$, with $[Y_{i} | S(x_{i})] ={\rm N}(\mu_{i} + S(x_{i}),\tau^2)$. 
Write $X^*$ for a set of unknown locations at which measurements $Y^*$ have been made,
$X=(\tilde{X},X^*)$ and $Y=(\tilde{Y},Y^*)$; the observed quantities are $\tilde{X}$ and $Y$. 
For any set ${\cal X}$ of 
points $x \in \mathbb{R}^2$ write 
$S({\cal X}) = \{S(x): x \in {\cal X}\}$; hence,
$S(X) = (S(\tilde{X}),S(X^*))$. Assume that $X$ and $S$ are stochastically 
independent. The joint distribution of $X$, $Y$ and $S$ is then
\begin{eqnarray}
\label{eq:joint_distr}
[X,Y,S] &=& [S]\: [X]\:[Y | S, X]  \nonumber \\
                        &=& [S]\: [\tilde{X}]\: [X^* | \tilde{X}]\:[\tilde{Y} | S(\tilde{X})] \:[Y^* | S(X^*)],
\end{eqnarray}
Our assumption that the sampling design is non-preferential allows a straightforward
marginalisation of (\ref{eq:joint_distr}) to give
\begin{eqnarray}
\label{eq:joint_distr2}
[X,Y] &=&  [\tilde{X}]\:  [X^* | \tilde{X}] \:[\tilde{Y} | X] \:[ Y^*| \tilde{Y},X] \nonumber \\
                     &=&  [\tilde{X}] \: [X^* | \tilde{X}] \: [\tilde{Y}|\tilde{X}] \:  
                                  [Y^*|\tilde{Y},X^*]\nonumber \\
                     & = & [\tilde{X}] \: [X^* | \tilde{X}] [Y|X]
\end{eqnarray}
where $[Y|X]$ is a multivariate Gaussian distribution with mean vector 
$\mu = (\mu_1,...,\mu_{n})$ and covariance matrix $\Sigma$ with
diagonal elements $\sigma^2+\tau^2$ and off-diagonal elements
 $\sigma^2 \rho(x_{i}, x_{j}; \phi)$.

\subsection{Sampling design}
\label{subsec:model_sampling_loc}
Depending on the problem under investigation, the set of sampling locations might be the result of a natural process, for example the locations of nests in a colony of birds. Alternatively, 
they might be obtained by using a random or regular lattice designs, as it is often the case for household surveys or agricultural field trials, respectively. Knowledge of the underlying process generating the sampling locations should then be incorporated into the specification
of the distribution $[\tilde{X}]$. We now briefly outline some approaches to this specification,
and propose a non-parametric approach that can be used when information on the underlying sampling process is limited.

One approach would be to model $[\tilde{X}]$ as an inhomogeneous Poisson process over the region of interest $A \subseteq \mathbb{R}^2$, with intensity
\begin{equation}
\label{eq:intensity_poisson_process}
\lambda(x) = d(x)^\top \beta\text{, }
\end{equation}
where $d(x)$ is a $p$-dimensional vector of spatial covariates, such as population
density in the case of a randomised household survey, and $\beta$ is the associated vector of regression coefficients. 

When no information on the sampling design is available a non-informative uniform distribution could
 be used, hence $\lambda(x) = \lambda$ for all $x \in A$. An alternative approach is to estimate
 the intensity $\lambda(x)$ from the data, using a kernel method. Let $x_{1}$ and $x_{2}$ denote the coordinates of the horizontal and vertical axes for a given point $x \in \mathbb{R}^2$. Then, the kernel density estimate of $\pi(x)$, i.e. the marginal density function of any component of $X$, based on the observed locations $X^*$ is given by
\begin{equation}
\label{eq:kernel_density}
\hat{\pi}(x) = \frac{1}{n}\sum_{i=1}^n K_{H}(x_{1}-x_{1i}; x_{2}-x_{2i}),
\end{equation}
where $K_{H}(\cdot; \cdot)$ is a bivariate kernel with symmetric and positive definite 2 by 2 smoothing matrix $H$. If we choose a Gaussian kernel, then $H$ is the variance matrix of
a bivariate Gaussian density and
$$
K_{H}(x_{1}-x_{1i}; x_{2}-x_{2i}) = \frac{1}{\sqrt{2 \pi}|H|^{1/2}}\exp\left\{-\frac{1}{2}(x-x_{i})^\top H^{-1}(x-x_{i})\right\}.
$$
The elements of $H$ can be  estimated by optimising an estimate of the mean-square-error
\citep*{breman1989}.
Alternatively, if we assume that $X$ is an independent random sample from a bivariate Gaussian distribution, the optimal $H$
in the sense of minimisng the integrated mean-square-error 
 is $ H = n^{-1/6} V$ \citep{lucy2002}, where $V$ is the sample covariance matrix. 

Finally, a common practice in geostatistical investigations is to choose locations that are more
regularly distributed over $A$ than would be a realisaton of a homogeneous Poisson process.  In
this case, a more appropriate prior for $[X]$ would be an inhibitory point
process \citep[pages 110-111]{diggle2013}; we give an example in Section \ref{subsec:inhibitory}.

\section{Computational details}
\label{sec:conditional_simulation}
\subsection{One unknown location}
\label{subsec:one_loc}
If $X^*$ consists of a single unknown location, numerical quadrature can be used for efficient computation of the predictive distribution $[X^*|\tilde{X},\tilde{Y},Y^*]$. Let $W = (w_{1},\ldots,w_{N})$ be a grid of spatial points in the region of interest $A$ and let $\pi(z)$ denote the density function of $Z$. It follows from \eqref{eq:joint_distr2} that
\begin{equation}
\label{eq:predictive_distr}
\pi(x^* | \tilde{x}, \tilde{y}, y^*) \propto \pi(x^* | \tilde{x})\pi(y^*| \tilde{y}, x^*). 
\end{equation}
By treating the Gaussian process $S$ 
as constant within each grid cell, we approximate the above density function by
\begin{equation}
\label{eq:predictive_approx}
\pi(x^* | \tilde{x}, \tilde{y}, y^*) \approx h(w^*) = \frac{\pi(w^* | \tilde{w}, \tilde{y}, y^*)}{\sum_{i=1}^N \pi(w_{i} | \tilde{w}, \tilde{y}, y^*)},
\end{equation}
where $w^*$ and the elements of $\tilde{w}$ are the grid points closest to $x^*$ and to 
the corresponding elements of
$\tilde{x}$, respectively. 
Summaries of the predictive distribution, such as mean, mode and component-wise median, can then be approximately computed through $h(\cdot)$. Additionally, high density regions of 
coverage $\alpha$ can also obtained as $$\left\{w_{i} \in W:l \in \mathbb{R},  \sum_{i: h(w_{i}) > l} h(w_{i}) = 1-\alpha\right\}.$$

\subsection{Multiple unknown locations}
\label{subsec:multiple_loc}
When there is more than one unknown location, 
the numerical solution is no longer feasible. Instead, we have developed
 an MCMC algorithm that takes account of the presence of 
the multiple modes that typically characterize the density 
function of $[X^*|\tilde{X},\tilde{Y},Y^*]$, each 
mode corresponding to a location where the absolute difference 
between a value of $Y^*$ associated with an unknown location
and an observed value of $\tilde{Y}$ is small. 
At each iteration of the MCMC,
and with a pre-specified probability, 
the algorithm proposes a draw from a mixture distribution with a mode centred on each observed location. \par
Let $n^*$ and $\tilde{n}$ denote the number of unknown and known locations, respectively; we propose the following Metropolis-Hastings algorithm to simulate 
from the target density given by \eqref{eq:predictive_distr}. 
\begin{enumerate}
\item Initialize $x_{\text{curr.}}^* = x_{0}^*$.
\item Propose a new value $x_{\text{prop.}}^*$ as follows. For $i=1,\ldots,n^*$:
\begin{itemize}
\item with probability $p$ perform a random walk by proposing a value from a bivariate Gaussian distribution with mean the $i$-th element of $x_{\text{curr.}}^*$ and covariance matrix $h_{1}^2I$, where $h_{1} > 0$ and $I$ is a 2 by 2 identity matrix;  
\item with probability $1-p$ sample a data point $x_{j}$ uniformly  from the set of observed locations and propose a value from a bivariate Gaussian distribution with mean $x_{j}$ and covariance matrix $h_{2}^2I$, where $h_{2} > 0$. 
\end{itemize}
\item Accept $x_{\text{prop.}}^*$ with probability 
$$\min\left\{1,\frac{\pi(x_{\text{prop.}}^* | \tilde{x}, \tilde{y}, y^*) q(x_{\text{curr.}}^*|x_{\text{prop.}}^*)}{\pi(x_{\text{curr.}}^* | \tilde{x}, \tilde{y}, y^*) q(x_{\text{prop.}}^*|x_{\text{curr.}}^*)}\right\},$$
where 
\begin{eqnarray}
q((w_{1},\ldots,w_{n^*})|(z_{1},\ldots,z_{n^*})) &=& \prod_{i=1}^{n^*} \Bigg[\frac{p}{h_{1}} \: f(\|w_{i}-z_{i}\|/h_{1})+ \nonumber\\
& &   \frac{1-p}{\tilde{n}h_{2}} \sum_{j=1}^{\tilde{n}}f(\|w_{i}-\tilde{x}_{j}\|/h_{2})\Bigg].
\label{eq:proposal}
\end{eqnarray}
In \ref{eq:proposal},
$f(\cdot)$ is the density function of a standard Gaussian distribution and $\|\cdot\|$ is the
 Euclidean distance.
\item Repeat 2 and 3 for the desired number of iterations. 
\end{enumerate}
In this algorithm, the standard deviations $h_{1}$, $h_{2}$ and $p$ should be tuned manually via pilot runs.

\section{Examples}
\label{sec:examples}

\subsection{Simulated data from a homogeneous Poisson process}
\label{subsec:example_simulation}

Our first example highlights the difficulty
 of summarizing the multi-modal distribution $[X^*|\tilde{X},\tilde{Y},Y^*]$ by a single measure of location.  We simulated two data sets of size $201$, using an isotropic exponential correlation function for the Guassian process $S(x)$. The parameter values were set as
 $\mu=0$, $\sigma^2=1$, $\tau^2=0.1$, and $\phi=0.1$ or $\phi=0.5$. 
From each of the two simulated data-sets we treated one of the 201 locations, chosen at random, as 
unknown. 
Locations were generated uniformly in the unit square, hence $\pi(x^* | \tilde{x}) = 1$ 
for $x^*$ lies in the unit square and 0 otherwise. \par
\begin{figure}
\begin{center}
\includegraphics[scale=0.7]{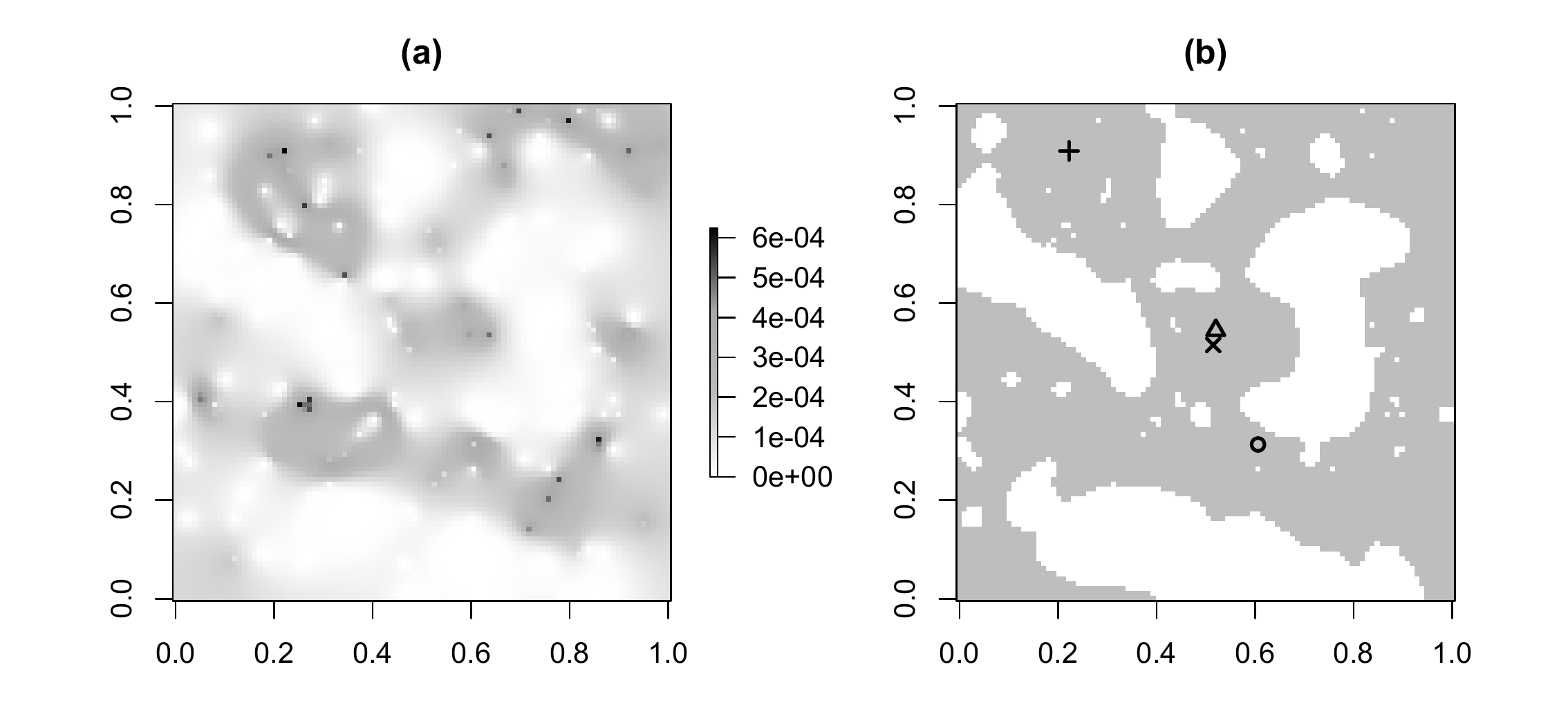} \\
\includegraphics[scale=0.7]{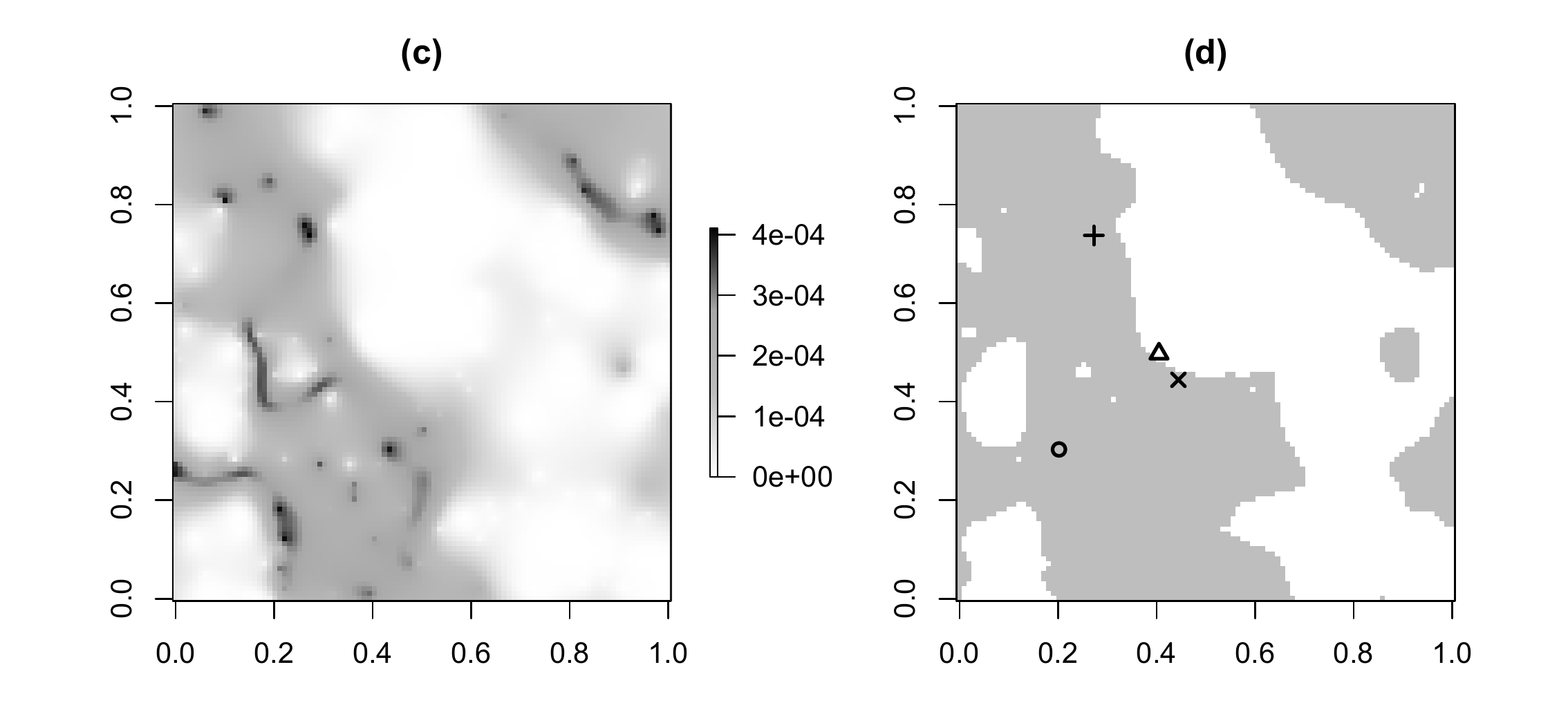}
\caption{(a)-(c) images of the predictive distribution of one missing locations, obtained from 200 hundred locations generated uniformly over the unit square and simulated data from a Guassian process with $\phi=0.1$ (a) and $\phi=0.5$ (c); (b)-(d) images of $95\%$ high density regions for the predictive distributions in (a) and (c), repsectively, showing the true location (o), the mean (\protect\marksymbol{triangle}{black}), the mode ($+$) and the componentwise median ($\times$).\label{fig:example_simulation}}
\end{center}
\end{figure}

Figures \ref{fig:example_simulation}(a)-(b) and \ref{fig:example_simulation}(c)-(d) show the results for $\phi=0.1$ and $\phi=0.5$, respectively. The multi-modality of the distribution is indicated by numerous black patches of high density, that become more spread when $\phi=0.5$. In Figure \ref{fig:example_simulation}(d), the mean lies outside the $95\%$ highest density region, whilst the mode, although not having that unpleasant feature, has only slightly larger
predictive density than other local modes.

\subsection{Simulated data from an inhibitory point process}
\label{subsec:inhibitory}

We now consider the case when $[X]$ is a simple sequential inhibition process on the unit square. 
Denote by  $\delta > 0$ 
 the minimum permissible distance between any two locations.
A sample from $X$ is obtained by a sequential sampling of points from a $100$ by $100$ regular lattice, where each new location $X_{i+1}$ given $\{X_{j}=x_{j}, j=1,\ldots,i\}$ is generated uniformly on the intersection of the unit square with $\{x \in [0,1]^2: \|x-x_{j}\|\geq \delta,j=1,\ldots,i\}$. \par
\begin{figure}
\begin{center}
\includegraphics[scale=0.61]{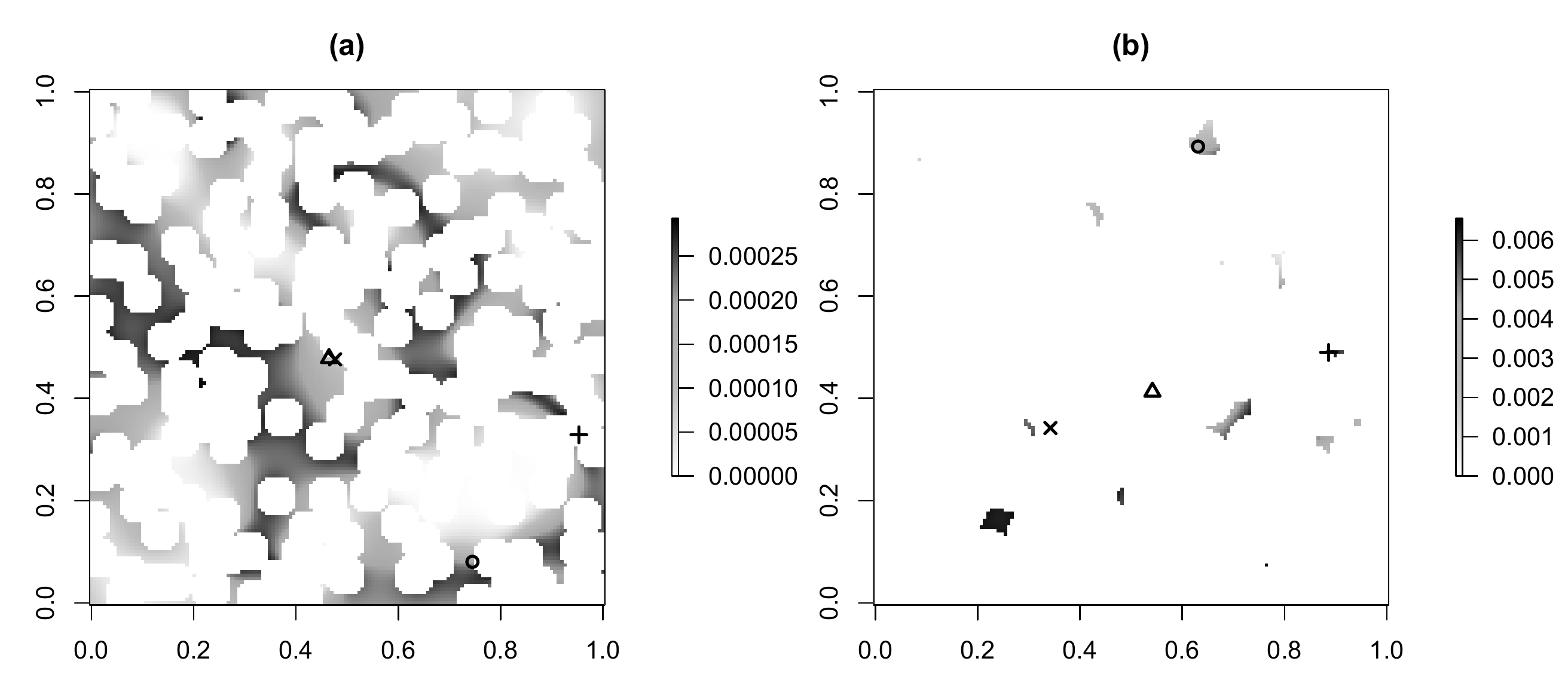} 
\caption{(a)-(b) images of the predictive distribution of one missing locations, obtained from 200 hundred locations generated from an inhibitory point process with minimum distance
between locations $\delta = 0.04$ (a) and $\delta = 0.06$ (b); the true location (o), the mean (\protect\marksymbol{triangle}{black}), the mode ($+$) and the component-wise median ($\times$) are also shown.\label{fig:example_inhibition}}
\end{center}
\end{figure}

We simulated two datasets of $201$ observations with model parameter values set as in 
the previous example but fixing $\phi=0.1$ and letting $\delta$ take values $0.04$ and $0.06$; again, only one location is treated as unknown. The resulting density of the predictive 
distribution is shown in Figure \ref{fig:example_inhibition}. As expected, the 
extent of the predictive distribution is reduced by imposition of the
minimum permissible distance $\delta$. In particular, for $\delta=0.06$ in Figure \ref{fig:example_inhibition}(b), only a few small, disjoint regions remain
as admissible for the missing location; note also that the mean and the median, in this specific example, lie outside the support of the predictive distribution.

\subsection{Rainfall data from Paran\'a State in Brazil}
\label{subsec:example_parana}
\begin{figure}
\begin{center}
\includegraphics[scale=0.55]{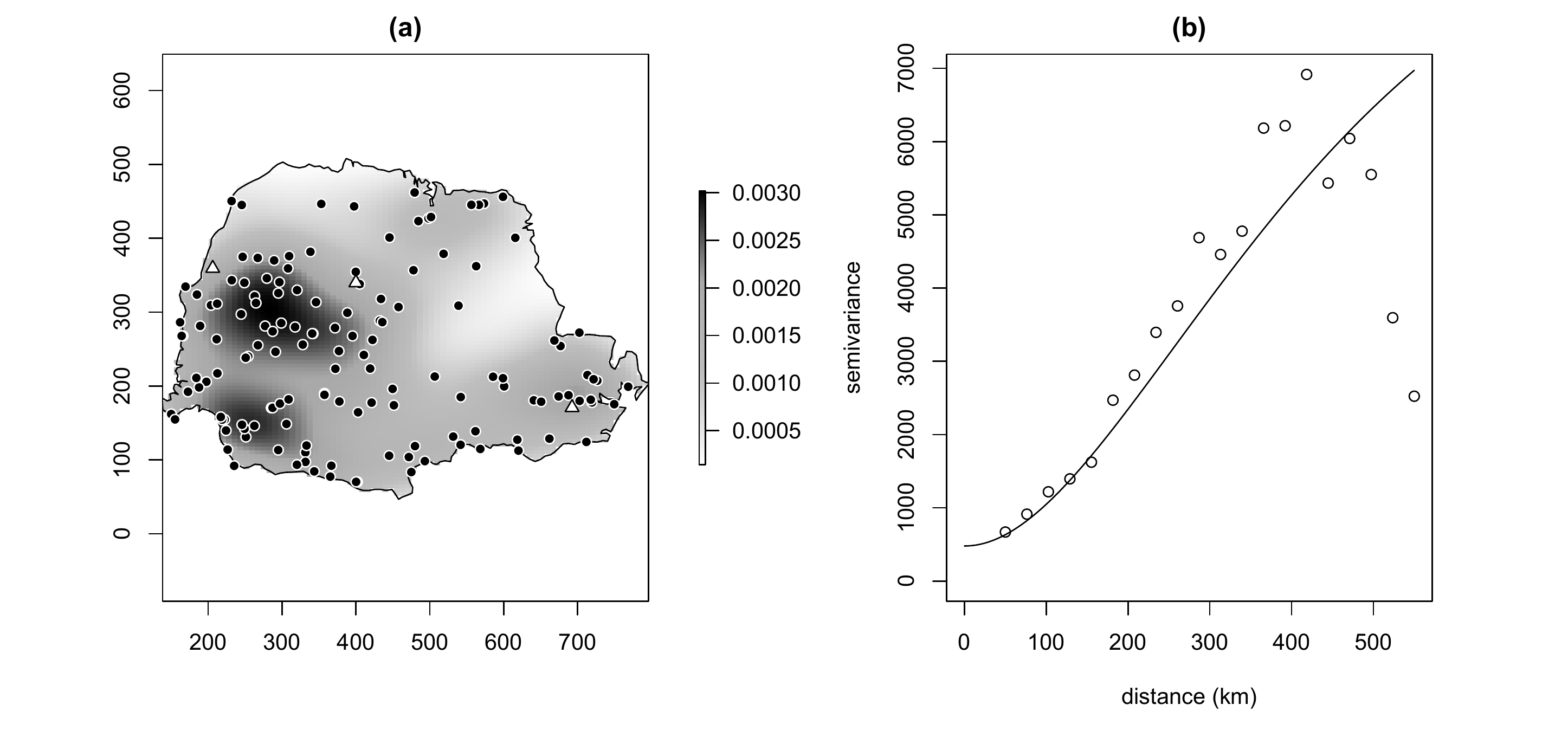}
\caption{(a) observed (solid cirlces) and missing (triangles) geographical coordinates of the recording station, with a kernel density estimate of $\pi(x^*)$ on the background; (b) empricical (points) and theoretical (solid line) semi-variogram.\label{fig:prior_parana}}
\includegraphics[scale=0.7]{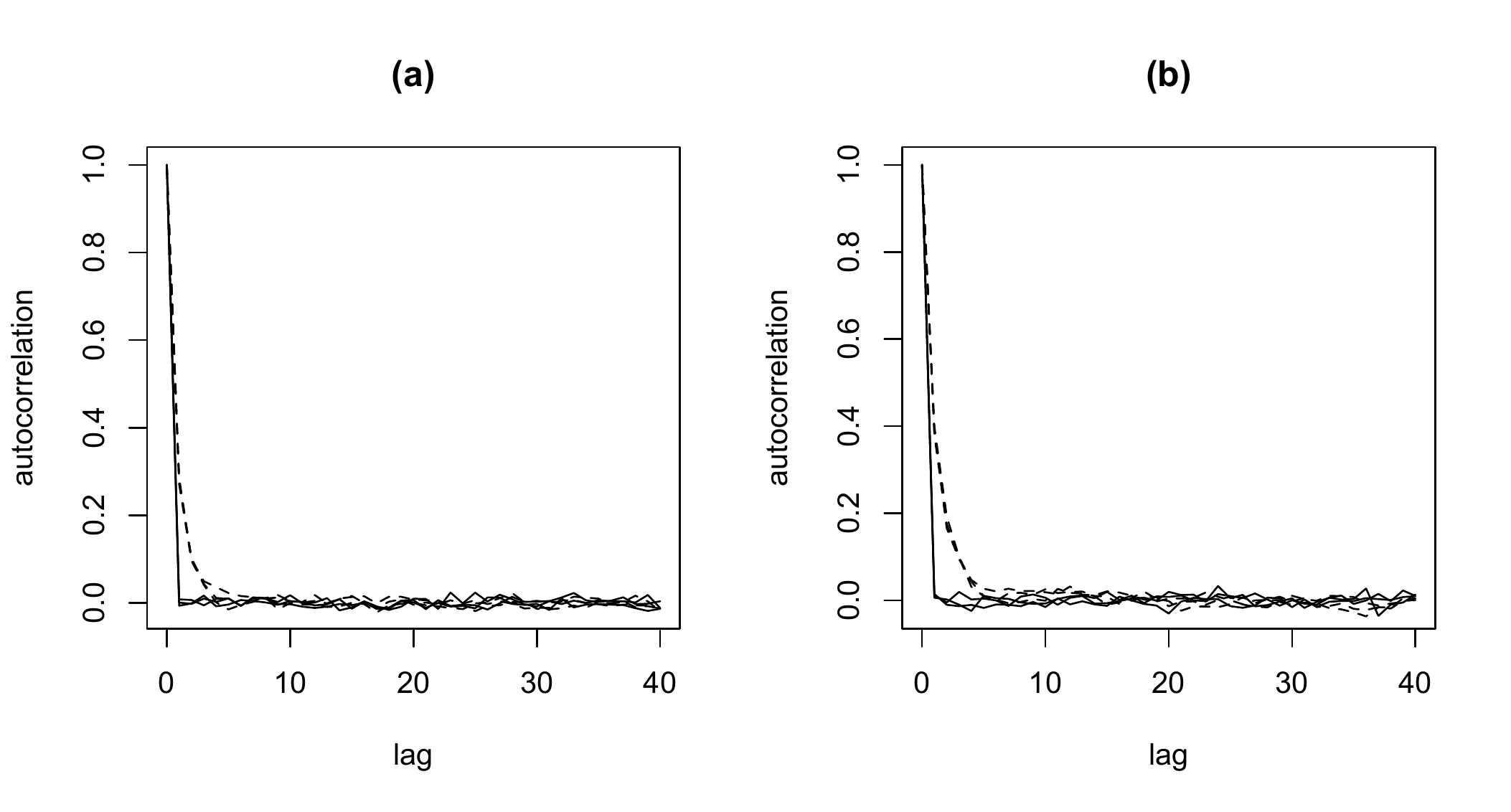}
\caption{autocorrelogram of the 10000 samples from $[X^*|X,Y,Y^*]$ for the horizontal (a) and (b) vertical coordinates, using a uniform distribution (solid lines) and a kernel density estimate (dashed lines) for $\pi(x^*)$.\label{fig:autocorrelogram}}
\end{center}
\end{figure}

We now consider a data set, previously analysed by \citet*{diggle2002}, on average rainfall over different years for the period May-June (dry-season) recorded in  143 recording stations throughout Paran\'a State, Brazil. Data locations are reported in Figure \ref{fig:prior_parana}, three of which, denoted by triangles, were randomly selected and treated as unknown. \par
Using the 140 observations with known locations, we first fitted the model $Y_{i} = \mu + S(x_{i}) + Z_{i}$, where $S(x_{i})$ is an isotropic Gaussian process with variance $\sigma^2$ and \citet{matern1986} correlation function
$$
\rho(u; \phi, \kappa) = \{2^{k-1}\Gamma(\kappa)\}^{-1}(u/\phi)^\kappa K_{\kappa}(u/\phi), u > 0,
$$ 
where $u$ is distance between two locations, $K_{\kappa}(\cdot)$ denotes the modified Bessel function of the second kind of order $\kappa >0$ and $\phi > 0$ is a scale parameter. The maximum likelihood estimates are: $\hat{\mu} = 251.539$, $\hat{\sigma}^2 = 9422.807$, $\hat{\tau}^2 = 479.074$, $\hat{\phi} = 200.004$ and $\hat{\kappa} = 1.913$. As also indicated by the resulting theoretical semi-variogram in Figure \ref{fig:prior_parana}(b), the data show evidence of long-range spatial correlation: the most distant pair of observations at 619.492 km from each other have an
estimated correlation of about 0.25. \par
In order to make inference on the three unknown locations, we model $[X^*]$ in two different ways: 1) $\pi(x^* | \tilde{x})$ is uniform over the square $[71.978, 846.652] \times [13.887, 518.441]$; 2) $\pi(x^* | \tilde{x})$ is estimated from the data using kernel density estimation as described in Section \ref{sec:model}. The resulting non-parametric estimate of $\pi(x^* | \tilde{x})$ is shown in Figure \ref{fig:prior_parana}(a); note that the boundaries of Paran\'a State play no part in
the analysis but are displayed only to add context.\par
\begin{figure}
\begin{center}
\includegraphics[scale=0.5]{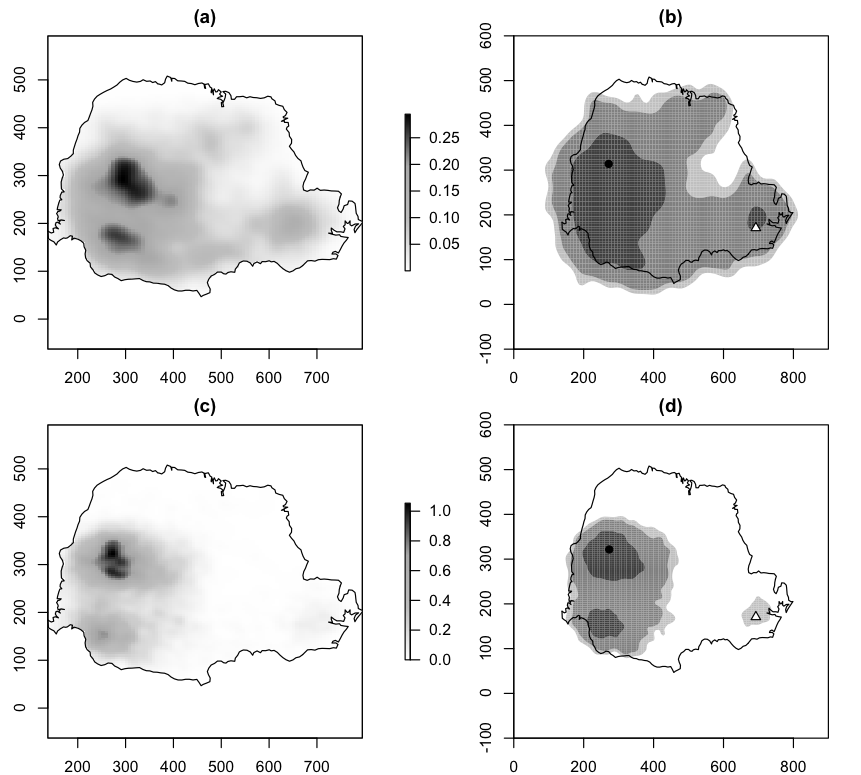}
\caption{(a),(c) images of the predictive distribution $\pi(x^* | \tilde{x}, \tilde{y}, y^*)$ for the missing location $x^* = (692.545, 170.875)$. (a) $\pi(x^*)$ uniform over the square $[71.978, 846.652] \times [13.887, 518.441]$. (c) kernel density estimate for $\pi(x^*)$; (b),(d)  $95\%$, $90\%$ and $50\%$ highest density regions for the predictive distributions in (a) and (c), respectively, indicating the mode (solid circle) and the true location $x^*$ (triangle).\label{fig:posterior_parana}}
\end{center}
\end{figure}

 
In both scenarios, 
using the MCMC algorithm described in Section \ref{sec:conditional_simulation}, we obtained 10000 samples from the predictive distribution $[X^*|X, Y, Y^*]$, iterating 110000 times and retaining every 10th sample after a burn-in of 10000 samples; we set $h_{1}=70$, $h_{2}=31$ and  $p=0.5$. The autocorrelogram plots for the vertical and horizontal coordinates in 
Figure \ref{fig:autocorrelogram} suggest rapid convergence of the MCMC algorithm. 
The resulting predictive distributions for one of the three missing location, specifically $x^* = (692.545, 170.875)$, are shown in Figure \ref{fig:posterior_parana} with corresponding high density regions. Panels (b) and (d) show that the high density regions obtained using a uniform distribution have a much wider extent than in the second case and are little informative on the possible positioning of $x^*$. However, when using the kernel density estimate $\hat{\pi}(x)$, the true location is only contained in the $95\%$ high density region; indeed, as indicated in Figure \ref{fig:prior_parana}(a), this point lies in a region where $\hat{\pi}(x)$ is relatively small.

\section{Discussion}
\label{sec:discussion}
In this paper we have proposed a methodology that allows to make inference on 
unknown locations of geostatistical data. We have developed efficient quadrature
and MCMC algorithms for 
sampling from the predictive distribution $[X^*|\tilde{X},\tilde{Y},Y^*]$ for single and
multiple unknown locations $X^*$, respectively, and applied these to two simulated data-sets
and to rainfall data from Paran\'a state, Brazil, with three unknown locations. 
In other examples not shown,
 we found that the MCMC algorithm continued to mix
 well when there are more unknown locations. \par
The conjunction of a sufficiently large data-set and many unknown locations would 
increase the computational burden of the 
 MCMC algorithm. 
A computationally more efficient, but approximate procedure would be to use numerical 
quadrature in turn for each of the missing locations. 
\par
As shown in Section \ref{subsec:example_simulation}, the 
distribution $[X^*|\tilde{X},\tilde{Y},Y^*]$ is often characterized by multiple modes and disjoint high density regions,
 in which case commonly used indices, such as the mean and the median, 
are misleading summaries. \par
We have shown that the use of a homogeneous Poisson process
prior  $[X^*]$ may result in a very diffuse predictive distribution
with widespread regions of high density.
The use of a kernel density estimate for $\pi(x^* | \tilde{x})$, based on the set of observed locations is useful when the empirical distribution of $\tilde{x}$ is spatially heterogeneous
and the conditional distribution of the unknown locations
 $X^*$ is expected to follow the same pattern. Conversely, an inhibitory process for
$[X^*|\tilde{X}]$ is  more appropriate when the context suggests that the complete set of
locations $X=(\tilde{X},X^*)$ is likely to show some degree of spatial
regularity.

We have not considered incorporation of observations with unknown
 locations into a likelihood for parameter estimation.
 However, this can be pursued by noticing that, starting from \eqref{eq:joint_distr2}, the likelihood is given by
$$
\pi(\tilde{y}, y^*) \propto \pi(\tilde{y} | \tilde{x}) \times 
           \int _{\mathbb{R}^{2n^{*}}} \pi(x^* | \tilde{x}) \pi(y^* | \tilde{y}, x^*) \: dx^*,
$$
where the first factor is the standard likelihood function obtained from data at
known locations. The second factor is an intractable integral of dimension
twice the number of unknown locations, $n^*$ say. As discussed in 
Section \ref{sec:conditional_simulation}, either
numerical quadrature or MCMC can be used to evaluate the integral according
to the value of $n^*$.  More efficient Monte Carlo techniques based on importance sampling or the EM algorithm could also be considered. For Bayesian inference, we would need additionally
to specify a prior for the model parameters. 
It would then be interesting to determine under what circumstances
 the incorporation of observations with missing locations into the model-fitting process
leads to materially better parameter estimates. We conjecture that the additional information will be small unless 
the prior for $[X^*|\tilde{X}]$ is highly informative.
 \par
Finally, we have assumed throughout that $[X]$, and hence 
$[X^*|\tilde{X}]$, is stochastically independent of $[S]$. 
Extensions of the  modelling framework that allow for
stochastic dependence between $X$ and the latent process $S$ would add to the predictability of unknown locations
$X^*$ but would also complicate the fitting of the model. One of a number of possibilities
would be  to model $X$ as a Cox Process with intensity stochastically dependent on
$S$, as in \citet*{diggle2010}.

\bibliographystyle{biometrika.bst}
\bibliography{biblio.bib}
\end{document}